\documentclass{aa}

\usepackage{comment}

\usepackage{natbib}
\bibpunct{(}{)}{;}{a}{}{,} 
\usepackage{amssymb}
\usepackage{url}
\usepackage{graphicx}
\usepackage{longtable}
\usepackage[usenames,dvipsnames]{color}
\newcommand{\submm}{submm}

\newcommand{\smgs}{{\submm} galaxies}
\newcommand{\smg}{{\submm} galaxy}

\newcommand{\micron}{\mbox{$\mu$m}}
\newcommand{\msun}{\mbox{${\rm M}_\odot$}}

\defcitealias{michalowski10smg}{M10} 	
\defcitealias{cirasuolo10}{C10} 
\defcitealias{dacunha08}{dC08}

\newcommand{\sunrise}{\textsc{sunrise}\xspace}
\newcommand{\gadgetthree}{\textsc{gadget-3}\xspace}
\newcommand{\mstar}{M_*}

\begin{document}

 \title{Determining the stellar masses of submillimetre galaxies: the critical importance of star formation histories}
 
\titlerunning{Determining the stellar masses of submillimetre galaxies}
\authorrunning{Micha{\l}owski et al.}

\author{Micha{\l}~J.~Micha{\l}owski\inst{\ref{inst:roe}}\thanks{{\tt mm@roe.ac.uk}} 
\and
Christopher~C.~Hayward\inst{\ref{inst:hay}} 
\and
James~S.~Dunlop\inst{\ref{inst:roe}} 
\and
Victoria~A.~Bruce\inst{\ref{inst:roe}}
\and
Michele~Cirasuolo\inst{\ref{inst:roe},\ref{inst:atc}}
\and
Fergus~Cullen\inst{\ref{inst:roe}}
\and
Lars Hernquist\inst{\ref{inst:cfa}}
	}

\institute{
SUPA\thanks{Scottish Universities Physics Alliance}, Institute for Astronomy, University of Edinburgh, Royal Observatory, Edinburgh, EH9 3HJ, UK 
\label{inst:roe}
\and
Heidelberger Institut f\"{u}r Theoretische Studien, Schloss-Wolfsbrunnenweg 35, D-69118 Heidelberg, Germany \label{inst:hay}
\and
UK Astronomy Technology Centre, Royal Observatory, Edinburgh EH9 3HJ \label{inst:atc}
\and
Harvard-Smithsonian Center for Astrophysics, 60 Garden Street, Cambridge, MA 02138, USA \label{inst:cfa}
}


\abstract{%
Submillimetre (submm) galaxies are among the most rapidly star-forming and most massive high-redshift galaxies; 
thus, their properties provide important constraints on galaxy evolution models. However, there is still a 
debate about their stellar masses and their nature in the context of the general galaxy population.
To test the reliability of their stellar mass determinations, we used a sample of simulated submm 
galaxies for which we created synthetic photometry. The photometry were used to derive their stellar masses 
via spectral energy distribution (SED) modelling, as is generally done with real observations. 
We used various SED codes ({\sc Grasil}, {\sc Magphys}, {\sc Hyperz} and {\sc LePhare}) and various alternative 
assumed star formation histories (SFHs). 
We found that the assumption of SFHs with two independent components enables the SED modelling codes to most 
accurately recover the true stellar masses of the simulated {\smgs}. 
Exponentially declining SFHs (tau models) lead to lower masses (albeit still formally consistent with the true stellar masses), while 
the assumption of single-burst SFHs results in a significant underestimation of the stellar masses.
Thus, we conclude that studies based on the higher masses inferred from fitting the SEDs of real {\smgs} with 
double SFHs are most likely to be correct, implying (as shown, for example, by Micha{\l}owski et al. 2012a) that  
{\smgs} lie on the high-mass end of the main sequence of star-forming galaxies. This conclusion 
appears robust to assumptions of whether or not {\smgs} are driven by major mergers, since the suite of simulated 
galaxies modelled here contains examples of both merging and isolated galaxies.
We identified discrepancies between the true and inferred stellar ages (rather than the dust attenuation) as the primary determinant of the success/failure of the mass recovery.
Regardless of the choice of SFH, the SED-derived stellar masses exhibit a factor of $\sim2$ scatter around the true value; this scatter is an inherent limitation of the SED modelling due to simplified assumptions (regarding, e.g., the SFH, detailed galaxy geometry and wavelength dependance of the dust attenuation). 
Finally, we found that the contribution of active galactic nuclei ($<60$\% at the $K$-band in these simulations) does not have any significant impact on the derived stellar masses.
}


\maketitle

\section{Introduction}
\label{sec:intro}

Galaxies selected with single-dish submillimetre (submm) telescopes ({\smgs}; see \citealt*{casey14} for a recent review) are among the most rapidly star-forming, most dusty and most massive galaxies at high redshifts, and they contribute significantly to the star formation activity  of the Universe \citep[e.g.,][]{hopkins10,michalowski10smg,wardlow11}. Hence, their number density and properties provide an important constraint on galaxy evolution models \citep{baugh05,fontanot07,coppin09,dave10,hayward13c,hayward13}.  Constraining the models with {\smgs} is only possible when the properties of a statistically significant sample are measured accurately. 

However, there is still a debate whether {\smgs} have high stellar masses \citep[$\simeq10^{11}$--$10^{12}\,\msun$;][]{borys05,dye08,hatsukade10,michalowski10smg, michalowski10smg4,ikarashi11,santini10,tamura10,yun12,johnson13,targett13,koprowski14,koprowski15,toft14,wiklind14} or significantly lower masses \citep[$\mbox{a few}\times10^{10}\,\msun$;][]{hainline11, wardlow11,bussmann12,casey13,rowlands14,simpson14}. In \citet{michalowski12mass}, we showed that this difference can be attributed to different assumptions in the spectral energy distribution (SED) modelling regarding the initial mass function (IMF), single-age stellar population (SSP) synthesis models and star formation histories (SFHs). When assuming consistent IMF and SSP models, we found that  `double' SFHs (with two independent components) result, on average, in stellar masses that are higher than those inferred by assuming `single' SFHs: 
exponentially declining SFHs (i.e., `tau models') yield masses that are $\sim0.1$--$0.3$ dex less, and single-burst SFHs yield masses that are $\sim0.3$--$0.4$ dex less.
However, it is difficult to constrain the SFHs of {\smgs} \citep[but see][]{dye08}, which hampers our ability to recognize which of these options represents the true properties of these galaxies. The dynamical masses derived from CO line widths provide upper limits on stellar masses, but in \citet{michalowski12mass} we showed that even the higher masses of {\smgs} derived using double SFHs are consistent with their dynamical masses.

Therefore, there is still a need to test the reliability of the derived stellar masses with an independent dataset. For that purpose, we use a set of hydrodynamical simulations for which we calculate synthetic SEDs using dust radiative transfer.
For each of the simulated galaxies, we have not only a set of synthetic photometric datapoints (which mimic the data available for real {\smgs}) to be used in the SED modelling but also the correct input set of parameters, which should ideally be recovered by the SED modelling. A related test of the estimates of star formation rates (SFR) derived from infrared luminosities is presented in \citet{hayward14b}, and a more-general analysis of how well
SED modelling recovers the parameters of simulated galaxies is presented in Hayward \& Smith (in preparation).

This paper is structured as follows.
In Sect.~\ref{sec:sim}, we describe the simulations from which we calculated the synthetic photometric data. 
Then, in Sect.~\ref{sec:sed}, we describe the SED modelling we applied to these data. We present our results in Sect.~\ref{sec:res}.
We discuss the implications of our results in 
Sect.~\ref{sec:discussion} and close with a summary in Sect.~\ref{sec:conclusion}.
We use a cosmological model with $H_0=70$ km s$^{-1}$ Mpc$^{-1}$,  $\Omega_\Lambda=0.7$ and $\Omega_m=0.3$.

\section{Simulations}
\label{sec:sim}

\begin{figure}
\begin{center}
\includegraphics[width=0.5\textwidth]{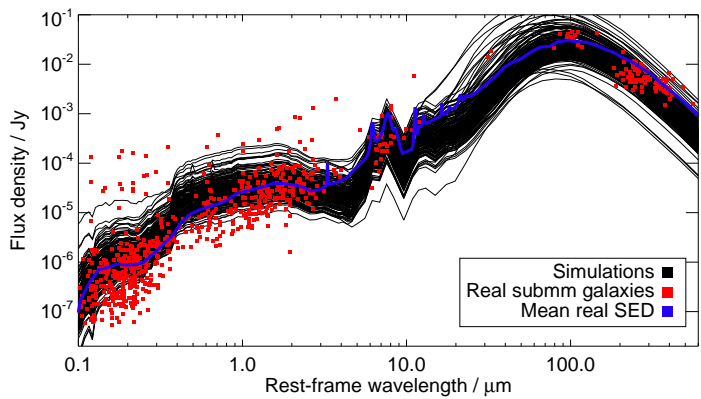}
\end{center}
\caption{Spectral energy distributions of the simulated {\smgs} ({\it solid black lines}) and the real {\smgs} ({\it red squares} for individual galaxies and {\it blue line} for the average model; both from the compilation of \citealt{michalowski10smg}), showing that the simulations correctly represent the real galaxies (no scaling of photometric datapoints has been applied).}
\label{fig:sed}
\end{figure}

\begin{figure*}
\begin{center}
\includegraphics[width=0.98\textwidth]{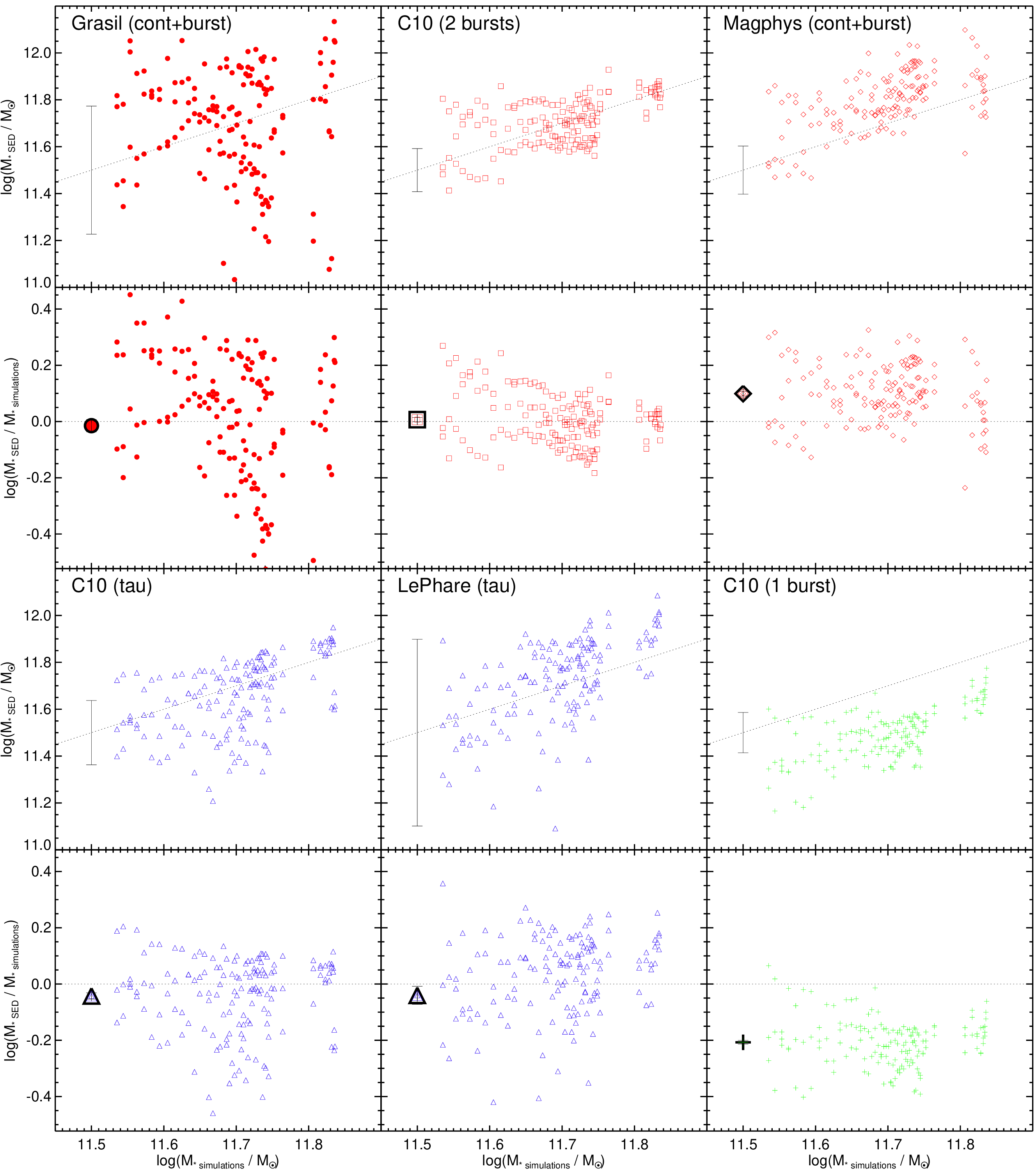}
\end{center}
\caption{Comparison of the stellar masses derived from the SED modelling and the true stellar masses in the simulation ({\it upper rows} for each model) and the ratio of these masses as a function of true stellar mass in the simulation ({\it lower rows}). {\it Dotted lines} indicate agreement between these masses. On each upper panels the {\it error bar} reflects the typical uncertainty in the stellar mass measurement. {\it Large symbols with error bars} at $\log(M_{\rm *\, simulations}/\msun)=11.5$ correspond to the mean ratios for each method. Models are colour-coded according to the assumed SFH: double ({\it red}), exponentially declining (tau, {\it blue}), and single-burst  ({\it green}). Except when a single-burst SFH is used, the SED modelling codes accurately recover the true stellar masses of
the simulated galaxies, albeit with a factor of $\sim 2$ scatter. When a single-burst SFH is assumed, the stellar masses are systematically underestimated by 0.2 dex.}
\label{fig:mcomp}
\end{figure*}

\begin{table*}
\caption{Stellar masses of simulated \smgs\ and the estimates recovered from the SED modelling. \label{tab:mstar}   }
\centering
\begin{tabular}{llccccc}
\hline\hline
Type\tablefootmark{a} & SFH\tablefootmark{b} & \multicolumn{2}{c}{$\log (M_* / M_\odot)$}  & \multicolumn{3}{c}{$\log(M_{\rm SED} / M_{\rm SIM})$}  \\
                     &                      & Mean & Median & Mean  & Median & Std. Dev. \\
\hline
Simulation & complex & $11.70 \pm0.01$ & $11.71 ^{+ 0.01 }_{- 0.01}$ & $\dots$ & $\cdots$\\
{\sc Grasil} & cont+burst & $11.68 \pm0.02$ & $11.73 ^{+ 0.02 }_{- 0.02}$ & $ -0.01 \pm0.02$ & $\phantom{-}0.04 ^{+ 0.02 }_{- 0.03}$ & $0.27$ \\
\citetalias{cirasuolo10} & double burst & $11.70 \pm0.01$ & $11.71 ^{+ 0.01 }_{- 0.01}$ & $\phantom{-}0.01 \pm0.01$ & $ -0.00 ^{+ 0.01 }_{- 0.00}$ & $0.09$ \\
{\sc Magphys} & cont+burst & $11.80 \pm0.01$ & $11.81 ^{+ 0.01 }_{- 0.01}$ & $\phantom{-}0.10 \pm0.01$ & $\phantom{-}0.10 ^{+ 0.01 }_{- 0.01}$ & $0.10$ \\
{\sc LePhare} & tau & $11.66 \pm0.03$ & $11.74 ^{+ 0.02 }_{- 0.02}$ & $ -0.04 \pm0.03$ & $\phantom{-}0.06 ^{+ 0.01 }_{- 0.01}$ & $0.40$ \\
\citetalias{cirasuolo10} & tau & $11.66 \pm0.01$ & $11.68 ^{+ 0.02 }_{- 0.02}$ & $ -0.04 \pm0.01$ & $ -0.01 ^{+ 0.02 }_{- 0.01}$ & $0.14$ \\
\citetalias{cirasuolo10} & single burst & $11.49 \pm0.01$ & $11.50 ^{+ 0.01 }_{- 0.01}$ & $ -0.21 \pm0.01$ & $ -0.20 ^{+ 0.01 }_{- 0.01}$ & $0.09$ \\
\hline
\end{tabular}
\tablefoot{ 
\tablefoottext{a}{Simulated values, or the SED fitting method: {\sc Grasil} \citep{silva98,iglesias07}, {\sc Magphys} \citep{dacunha08}, the method of \citet{cirasuolo10}, and {\sc LePhare} \citep{arnouts99,ilbert06}.}
\tablefoottext{b}{Assumed star formation history being either continuous SFH with a burst (cont+burst), double burst, single burst, or exponentially declining (tau).}
}
\end{table*}

We used the results of the galaxy merger simulation presented in \citet{hayward11b}. The galaxy evolution was first simulated
using the $N$-body/smoothed particle hydrodynamics (SPH)\footnote{Although the standard formulation of SPH was used for this work, this should not be cause for
concern because the results of idealized galaxy merger simulations are relatively insensitive to the inaccuracies inherent in this technique \citep{hayward14}.}
code \gadgetthree \citep{springel05gadget}, which includes the effects of gas cooling, star formation,
supernova feedback, black hole accretion and active galactic nucleus (AGN) feedback. Then, the 3-D polychromatic Monte Carlo dust radiative transfer
code \sunrise \citep*{jonsson06,jonsson10} was used to calculate ultraviolet-mm SEDs of the simulated galaxies at various snapshots in
time from three different viewing angles. 
The attenuation curve that results from the radiative transfer calculations depends on the relative distribution of stellar and AGN sources and dust and thus cannot (necessarily) be simply represented as a single attenuation law.
The hydrodynamical simulations, radiative transfer calculations, and details of the specific merger simulation used
are discussed comprehensively in \citet{hayward11b,hayward12,hayward13}.
The {\smgs} simulated in this manner have properties that agree well with many properties of real {\smgs}, including
the typical SED \citep{narayanan10}, CO excitation ladders \citep{narayanan09}, relation between effective dust temperature and luminosity \citep{hayward12,narayanan10b}, relation between 850--\micron~flux
($S_{850}$) and stellar mass \citep[$\mstar$;][]{michalowski12mass,hayward13b,davies13}, and redshift distribution and number counts \citep{hayward13c,hayward13,karim13,weiss13,chen13b}. 
As shown in Fig.~\ref{fig:sed}, 
the shapes of the spectral energy distributions of most of the real {\smgs} are within the range found in simulations
(though we note that the range of optical fluxes of real {\smgs} is somewhat larger than that in simulations, which implies that particularly dusty or optically-bright {\smgs} are not represented in our simulations).
Thus, it is reasonable to use the simulated {\smgs} to test the methods used to infer the stellar masses of real {\smgs}.

To generate the mock photometry, we started with the rest-frame SEDs of the simulated galaxies for all 140 snapshots viewed from three different viewing angles.
We redshifted each SED by a random redshift drawn from a log-normal distribution with $\bar{z} = 2.6$ and $\sigma = 0.2$ to mimic the redshift distribution
of real {\smgs} \citep[e.g.,][]{yun12,michalowski12}. If $S_{850} > 3$ mJy, the SED was added to the mock {\smg} catalog. This process yielded photometry for 153  
mock {\smgs}.
For the SED codes with only stellar emission implemented ({\sc Hyperz} and {\sc LePhare}\footnote{The latest version of {\sc LePhare} can utilise far-IR data, but this version was not used in this work.}),
we generated photometry for the $UBVRIzJHK$ and {\it Spitzer} channel 1 and 2 bands by convolving with the
appropriate filter response curves.
For codes that treat dust emission ({\sc Grasil} and {\sc Magphys}), we added synthetic photometry at $24$, $70$, $100$, $160$, $250$, $350$, $500$, $850$, $1100$ and $2000\,\mu$m. 

The experiment was performed in a blind manner: CCH performed the process described above and provided MJM, VAB, MC, and FC with only the photometry and redshifts for the 153
mock {\smgs} (i.e., the true stellar masses were not provided). 
The mock photometry was used to infer the stellar masses (Sect.~\ref{sec:sed}), and then
we compared the inferred and true masses. The fitting codes were not modified or tuned in any manner during this process.

\begin{figure*}
\begin{center}
\includegraphics[width=\textwidth]{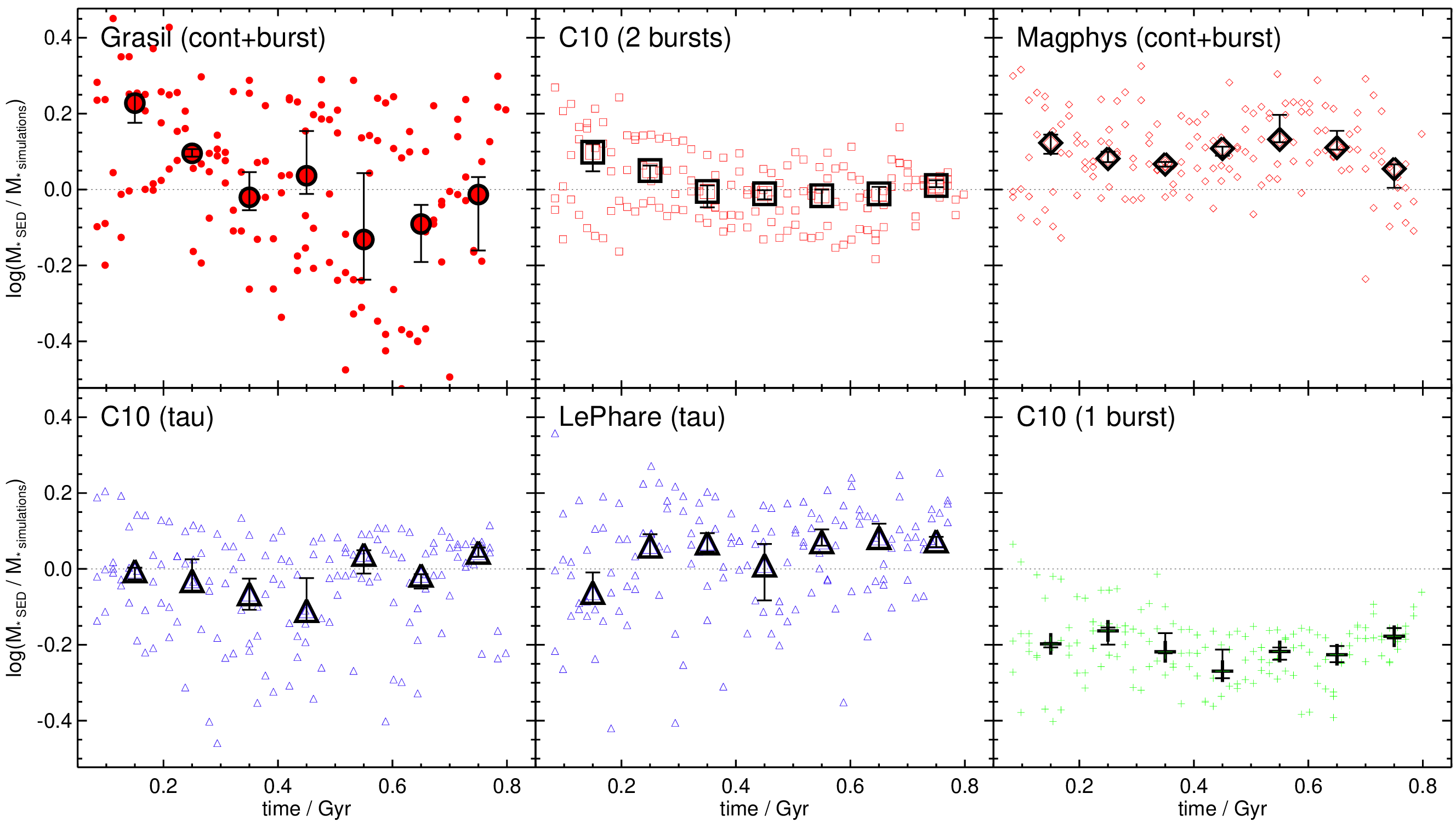}
\end{center}
\caption{Ratio of the stellar mass derived from the SED modelling to the true stellar mass in the simulation as a function of  the time elapsed in the simulation (i.e., evolutionary stage of the merger).
{\it Dotted lines} indicate agreement between these masses.  
{\it Large symbols} with error bars represent the medians in the time bins.
In all cases, there is no trend with the simulation time.}
\label{fig:mtime}
\end{figure*}

\begin{figure*}
\begin{center}
\includegraphics[width=\textwidth]{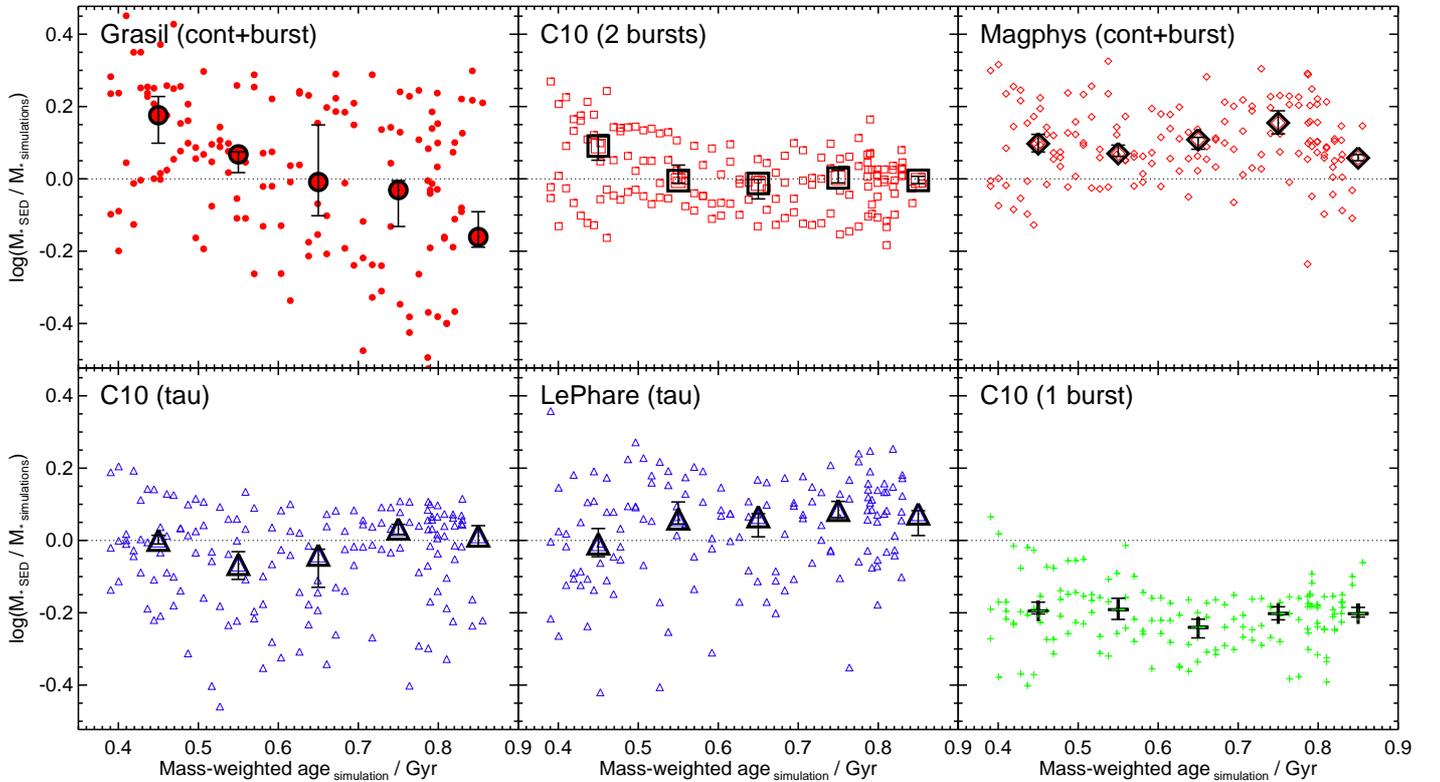}
\end{center}
\caption{Ratio of the stellar mass derived from the SED modelling to the true stellar mass in the simulation as a function of  the mass-weighted age of the stellar population in the simulations.
{\it Dotted lines} indicate agreement between these masses.  
{\it Large symbols} with error bars represent the medians in the age bins.
In all cases, there is no trend with the stellar age.}
\label{fig:mage}
\end{figure*}

\begin{figure*}
\begin{center}
\includegraphics[width=\textwidth]{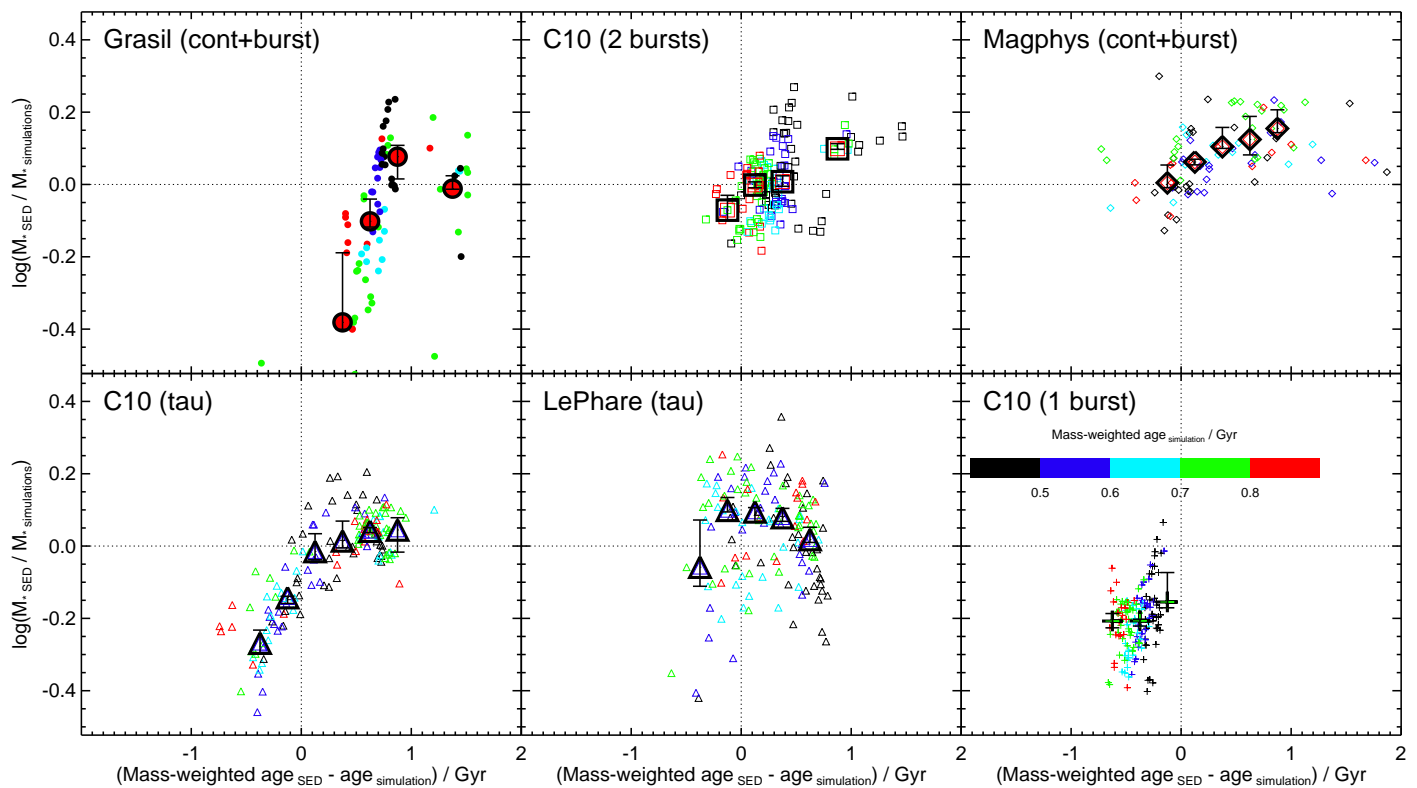}
\end{center}
\caption{Ratio of the stellar mass derived from the SED modelling to the true stellar mass in the simulation as a function of  the difference of the mass-weighted age of the stellar population recovered in the SED modelling and in the simulations.
{\it Dotted lines} indicate agreement between these masses and ages.  
{\it Large symbols} with error bars represent the medians in the age difference bins.
All points are colour-coded by the mass-weighted age in the simulations (colour bar).
For most models the recovered-to-true stellar mass ratio is correlated with the difference between the recovered and true mass-weighted stellar age.
}
\label{fig:magediff}
\end{figure*}

\begin{figure*}
\begin{center}
\includegraphics[width=\textwidth]{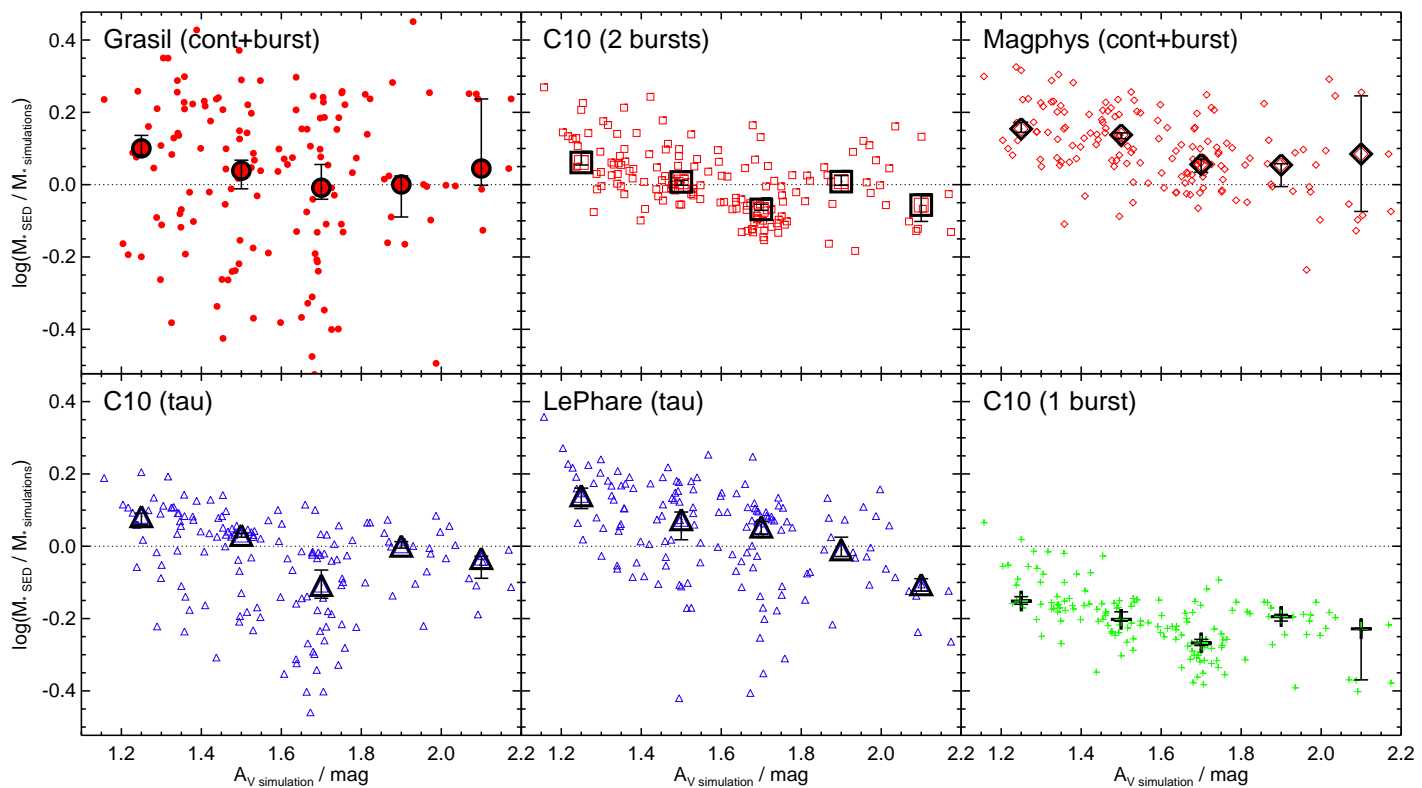}
\end{center}
\caption{Ratio of the stellar masses derived from the SED modelling and the true stellar masses in the simulation as a function of optical $V$-band dust attenuation in the simulation. {\it Dotted lines} indicate agreement between these masses. 
{\it Large symbols} with error bars represent the  medians in the $A_V$ bins. When a single-component (i.e., tau or single-burst) SFH is assumed, the recovered-to-true stellar mass ratio tends to decrease with increasing $A_V$.}
\label{fig:mav}
\end{figure*}

\begin{figure*}
\begin{center}
\includegraphics[width=\textwidth]{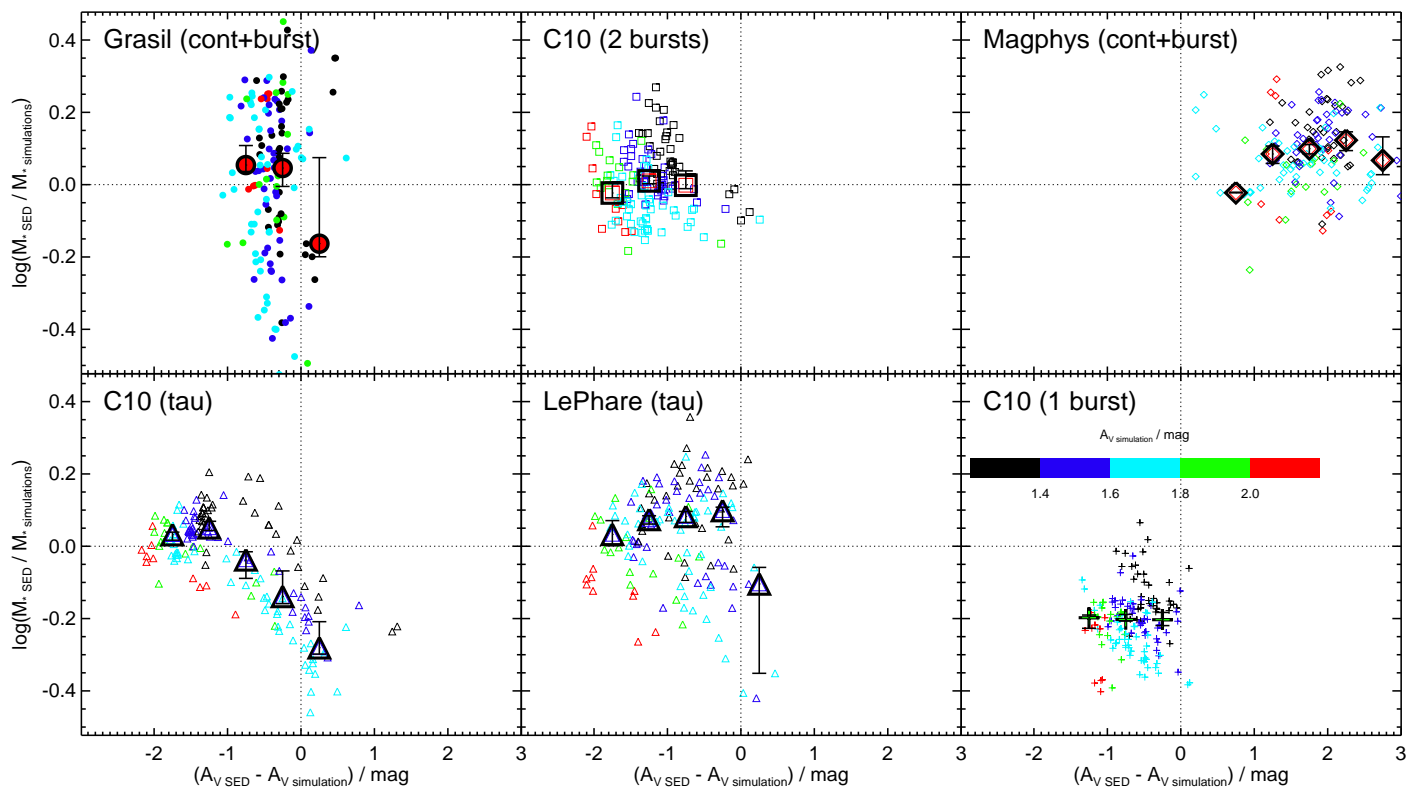}
\end{center}
\caption{Ratio of the stellar masses derived from the SED modelling and the true stellar masses in the simulation as a function of the difference between the optical $V$-band dust attenuation recovered in the SED modelling and in the simulation. {\it Dotted lines} indicate agreement between these masses and attenuations. 
{\it Large symbols} with error bars represent the  medians in the $A_V$ difference bins. 
All points are colour-coded by the dust attenuation in the simulations (colour bar).
These plots indicate that for all models except the C10 (tau) model, discrepancies between the true and recovered $A_V$ values are not the primary cause of inaccurate stellar mass recovery.
}
\label{fig:mavdiff}
\end{figure*}

\begin{figure*}
\begin{center}
\includegraphics[width=\textwidth]{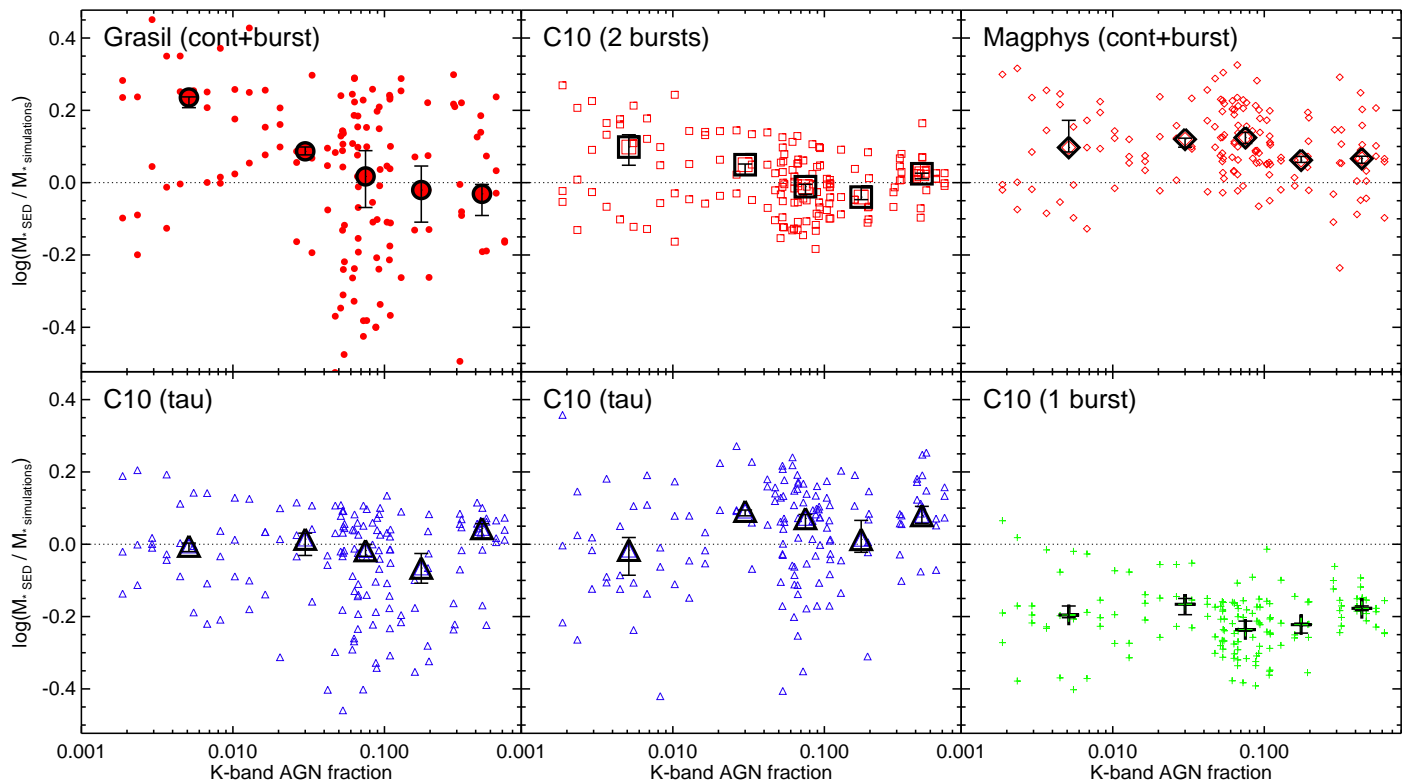}	
\end{center}
\caption{Ratio of the stellar masses derived from the SED modelling and the true stellar masses in the simulation as a function of the intrinsic $K$-band AGN fraction. {\it Dotted lines} indicate agreement between these masses.
{\it Large symbols} with error bars represent the medians in the AGN fraction bins. In all cases, there is no significant trend with the $K$-band AGN fraction.
}
\label{fig:agn}
\end{figure*}

\begin{figure*}
\begin{center}
\includegraphics[width=\textwidth]{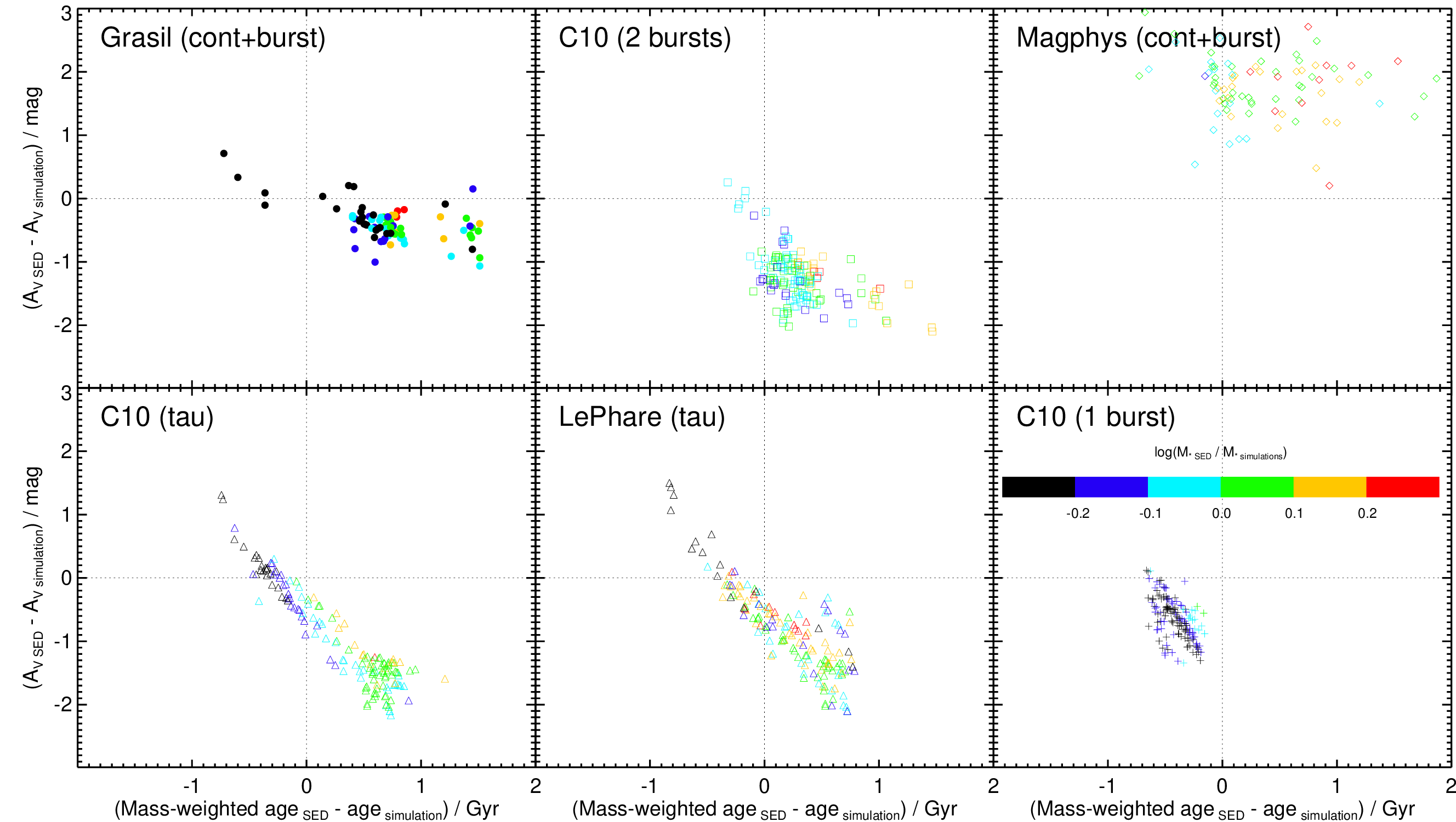}
\end{center}
\caption{Difference between the optical $V$-band dust attenuation recovered in the SED modelling and in the simulation as a function of  the difference of mass-weighted age of the stellar population recovered in the SED modelling and in the simulations. {\it Dotted lines} indicate agreement between these ages and attenuations. 
All points are colour-coded by the ratio of the recovered and simulated stellar mass (colour bar). 
The strong correlations observed for the tau models indicate that these fits are affected by the degeneracy between stellar age and dust reddening. Overall, this plot reinforces the conclusion that discrepancies between the true and recovered ages are the primary cause of inaccurate mass recovery.
}
\label{fig:avdiffagediff}
\end{figure*}

\section{SED fitting}
\label{sec:sed}

Both the simulations and all of the fitting codes use similar stellar population synthesis models.
 Moreover, we consistently use the definition of stellar mass as the total gas mass converted into stars during the galaxy's past evolution (i.e., the integral of the SFH). No stellar mass loss or decrease in the stellar mass due to dying stars is taken into account in either the simulations or the SED modelling. Throughout we assume the \citet{chabrier03} IMF.

\subsection{{\sc Grasil}}

We applied the SED fitting method detailed in \citet[][see therein a discussion of the derivation of galaxy properties and typical uncertainties]{michalowski08,michalowski09,michalowski10smg,michalowski10smg4} which is based on 35\,000 templates from the library of \citet{iglesias07} plus some templates of \citet{silva98} and \citet{michalowski08}, all of which were developed using {\sc Grasil}\footnote{\url{http://adlibitum.oats.inaf.it/silva/grasil/grasil.html}} \citep{silva98}. They are based on numerical calculations of radiative transfer within a galaxy, which is assumed to be a triaxial axisymmetric system with diffuse dust and dense molecular clouds, in which stars are born.

The templates cover a broad range of galaxy properties from quiescent to starburst and span an $A_V$ range from $0$ to $5.5$ mag. The extinction curve \citep[Fig.~3 of][]{silva98} is derived from the modified dust grain size distribution of \citet{draine84}.
The star formation histories are assumed to be a smooth Schmidt-type law \citep[i.e., the SFR is proportional to the gas mass to some power; see][for details]{silva98} with a starburst (if any) on top of that, starting $50$ Myr before the time at which the SED is computed. There are seven free parameters in the library of \citet{iglesias07}: the normalization of the Schmidt-type law, the timescale of the mass infall, the intensity of the starburst, the timescale for molecular cloud destruction, the optical depth of the molecular clouds, the age of the galaxy and the inclination of the disk with respect to the observer.

The code assumes the \citet{salpeter}   IMF; thus, the stellar masses were divided by 1.8 to convert to the \citet{chabrier03} IMF.

\subsection{{\sc Magphys}}

We also used {\sc Magphys}\footnote{\url{www.iap.fr/magphys}} \citep[Multi-wavelength Analysis of Galaxy Physical Properties;][]{dacunha08}, which is an empirical, physically-motived SED modelling code that is based on the energy balance between the energy absorbed by dust and that re-emitted in the infrared. We used the \citet{bruzualcharlot03} stellar population models and adopted the \citet{chabrier03} IMF. 

Similarly to {\sc Grasil}, in {\sc Magphys}, two dust media are assumed: a diffuse interstellar medium (ISM) and dense stellar birth clouds. Four dust components are taken into account: cold dust ($15$--$25$ K), warm dust ($30$-$60$ K), hot dust ($130$--$250$ K) and polycyclic aromatic hydrocarbons (PAHs). A simple power-law attenuation law is assumed.

\subsection{{\sc Hyperz} (C10)}

As in \citet{michalowski12mass}, we  also used the method presented in \citet[][hereafter C10]{cirasuolo07,cirasuolo10}, which is
based on the {\sc Hyperz} package \citep{bolzonella00}. We
utilised the \citet{bruzualcharlot03} models and assumed three different forms of SFH: 
an instantaneous burst of star formation, an exponentially declining SFH (tau model), and a double-component model composed of two instantaneous
bursts with different ages and (independent) dust attenuations ($0<A_V<6$ mag). In the double-component model, the age of the young component was varied between 
$50$\,Myr and $1.5$\,Gyr. The old component was allowed to contribute $0$--$100$\% of the near-IR emission, and its age was varied over the range $1$--$6$\,Gyr. Solar metallicity and the attenuation curve of \citet{calzetti00} were assumed. 
  These models were run using the Chabrier IMF.

\subsection{{\sc LePhare}}

Finally, we used {\sc LePhare}\footnote{\url{www.cfht.hawaii.edu/~arnouts/lephare.html}} \citep[PHotometric Analysis for Redshift Estimate;][]{arnouts99,ilbert06}. This code also uses the   \citet{bruzualcharlot03} models and the Chabrier IMF. Tau SFHs with $\tau$ in the range $0.03$--$30$ Gyr and age between 50 Myr and the age of the Universe at the given redshift were used.  The attenuation curve of \citet{calzetti00} was adopted, and the allowed $A_V$ range was $0$--$4$ mag. Solar metallicity was assumed.

\section{Results}
\label{sec:res}

The stellar masses inferred from the SED modelling are compared with the true masses in the simulation in the first and third rows of Fig.~\ref{fig:mcomp}. The ratios of the SED-inferred stellar masses to the true masses are shown in the second and forth rows of this figure. The mean and median masses and recovered-to-true stellar mass ratios are presented in Table~\ref{tab:mstar}. 
The recovered-to-true stellar mass ratios are also shown as a function of the time elapsed in the simulation (Fig.~\ref{fig:mtime}), mass-weighted age of the stellar populations of the simulated {\smgs} (Fig.~\ref{fig:mage}), the difference between the recovered and simulated age (Fig.~\ref{fig:magediff}), the mean optical dust attenuation in the simulation (Fig.~\ref{fig:mav}), the difference between the recovered and simulated attenuation (Fig.~\ref{fig:mavdiff}), and the $K$-band AGN fraction in the simulation (Fig.~\ref{fig:agn}). Finally, the difference in age is shown as a function of the difference in dust attenuation (colour-coded by the mass ratio) in Fig.~\ref{fig:avdiffagediff}.

For the simulated {\smgs}, we found similar trends as in \citet{michalowski12mass}: the stellar masses that result from the `double' SFH assumption are greater than those produced by the tau-model (by $\sim0.05$--$0.15$ dex) and single-burst (by $\sim0.2$--$0.3$ dex) SFHs.

The  SED models with double SFHs ({\sc Grasil} and C10~2c) accurately recover the correct simulated stellar masses. On average, the true and recovered stellar mass values differ by less than  0.01 dex.  Moreover, the double-burst model (C10~2c) results in the lowest scatter around the true stellar mass (last column of Table~\ref{tab:mstar}). We stress that the SFHs in the simulations result from the hydrodynamical calculations and do not exhibit shapes which directly resemble the double SFHs assumed in the SED modelling. In fact, the ages of the stellar population in younger models are not grossly underestimated by the single-burst model (Fig.~\ref{fig:magediff}), and some of the simulated {\smgs} are better fitted by single SFHs (see below).

 In contrast, {\sc Magphys} 
results in stellar masses that are on average $0.1$ dex higher than the true values in the simulations. 
However, we note that \citet{rowlands14} showed that using the standard {\sc Magphys} priors (which were applied here) results in stellar masses that are on average $\sim0.17$ dex higher than those that result when a different set of priors
(that allow higher dust attenuation levels and are thus more appropriate for {\smgs}) are used. Hence, it is likely that {\sc Magphys} would recover the stellar masses of the simulated {\smgs} more accurately if those priors
were used.

The assumption of exponentially declining (tau) SFHs (C10~tau and {\sc LePhare}) results in systematically lower stellar masses compared with double SFHs,
but the offset from the correct value is small on average ($\sim 0.05$ dex) and consistent with zero within the errors. There is no trend with simulation time (i.e., evolutionary stage of the merger) or with the AGN fraction.
However, there is an anti-correlation between $M_{\rm *\, SED\, tau}/M_{\rm *\, simulations}$ and the true value for the optical $V$-band dust attenuation, $A_V$, in the simulation.

Finally, single-burst SFHs (C10~1c) result in stellar masses that are systematically underestimated by $0.2$ dex. No trend with time (or AGN fraction) is found, in the sense that the discrepancy is a constant offset that
does not change with the evolutionary stage of the simulated {\smgs}. 

The reduced $\chi^2$ resulting from all models were similar (ranging from 1 to 6), which makes it difficult to choose the best fitting SFHs based on the goodness-of-fit. Reassuringly however, the models which reproduce the stellar masses correctly (C10~2c) resulted in lower reduced $\chi^2$ than single-burst models in $80$\% of cases and than tau models in $95$\% of cases.

\section{Discussion}
\label{sec:discussion}

It is worthwhile to consider which assumptions and uncertainties of the SED modelling methods used to infer stellar mass we can actually test. To perform dust radiative transfer
on the hydrodynamical simulations, it is necessary to assume an IMF (which may differ in SMGs from that observed locally; e.g., \citealt{baugh05,dave10,narayanan12b,narayanan13};
but cf. \citealt{hayward13}) and SSP templates. Consequently,
we obviously cannot check the validity of the IMF or SSP templates used in the SED fitting codes. Instead, we have ensured that the IMF and SSP templates used for
the dust radiative transfer and the SED fitting are consistent, which implies that any discrepancies between the true and inferred stellar mass values are not
caused by differences in the IMF or SSP templates. Uncertainties due to the IMF and SSPs (see, e.g., \citealt{michalowski12mass} for discussion) must therefore be added to the
uncertainties demonstrated here.

The aspects of the SED modelling that we \textit{can} test are primarily the assumptions made regarding SFH, AGN contamination,
dust attenuation and galaxy geometry. At the beginning of the simulations, there is a pre-existing stellar population in each of the progenitor disk galaxies;
the subsequent SFH of the simulation is a consequence of the hydrodynamics and the star formation model used. This simulated SFH is much more complex than a simple,
smoothly varying function
(see \citealt{hayward11b}); thus, we can measure the impact of assuming simple SFHs on derived stellar masses. 
In the simulations, the AGN emission
is determined by the accretion rate of the black hole particles, which depends on the hydrodynamics. Furthermore, it is attenuated by galaxy-scale dust.
Thus, the AGN can provide a non-trivial contribution to the mock SEDs (see \citealt{snyder13} for a detailed study), which can potentially significantly affect the stellar mass estimates;
our experiment tests how important such AGN contamination is (note that no SED code that we used accounts for AGN emission).
Finally, in the simulation, the input emission (from stars and AGN) varies with both (3-D) space and time, as does the dust content. Consequently,
the effective galaxy-scale dust attenuation law is not guaranteed to have a simple universal form, which is often assumed by SED modelling codes, and
the geometry is not necessarily well-described by the simple spherical or axisymmetric forms used in the SED modelling codes. Thus, our experiment can elucidate the uncertainty and/or biases
caused by the assumptions about dust attenuation and galaxy geometry that are made by the SED modelling codes.

As shown in Sect.~\ref{sec:res}, the SED models with double SFHs recover the correct simulated masses on average. This implies that the simplified assumptions made during the SED modelling regarding the analytical shape of the SFHs, the attenuation curve, and the galaxy geometry do not introduce any systematic offset in the resulting stellar masses (albeit they may explain some of the scatter). Thus, the stellar masses of real {\smgs} derived using these methods can be trusted. This is consistent with \citet{pacifici12}, who found that the stellar mass-to-light ratio of simulated galaxies with $\log(M_*/\msun)=9.5$--$11.5$ can be accurately recovered from broad-band photometry.  However, the apparent scatter of $\sim0.3$ dex visible in Fig.~\ref{fig:mcomp} illustrates that these simplified assumptions make the stellar masses derived for individual {\smgs} uncertain by a factor of two.

Tau models were shown by  \citet[][their Fig.~4]{mitchell13} to give systematically low stellar masses (by $\sim0.13$ dex) for starburst galaxies, which they defined as galaxies with higher SFRs in the burst component than in the smooth component. This underestimate is greater than the underestimation we find here for {\smgs} ($\sim0.05$ dex). However, their selection tends to yield more outlying objects compared with our {\smg} selection, which may also include galaxies with burst components lower than $50$\% \citep[only $\sim10$\% of stars in real {\smgs} were formed in recent bursts;][]{michalowski10smg,dye08}. Thus, in our case, the tau models result in much lower offsets from the true values. Similarly, \citet{wuyts09} and \citet{sobral14} showed that stellar masses are underestimated by $0.06$--$0.13$~dex for simulated galaxies with $M_*>1.4\times10^{10}\,\msun$ and by $0$--$0.06$~dex for $z\sim0.8$--$2.2$ H$\alpha$ emitters, respectively, when using tau or constant SFH models. Finally, \citet{simha14} found that tau models result in stellar masses that are $~0.05$~dex too low for $z<1$ galaxies with masses $M_*>10^{10}\,\msun$.

The inability of single-burst SFHs to recover the true stellar masses of the simulated {\smgs} occurs because their SFHs cannot be described by a simple, single-component form, and their past star formation contributes significantly to their total stellar masses, as is the case for real {\smgs} \citep[e.g.,][]{michalowski10smg}. Naturally, real SMGs are unlikely to have formed all their stars during a single epoch. We demonstrate here that assuming a single-burst SFH leads to the underestimation of stellar masses, unlike other simplifying assumptions, such as a symmetric geometry and smoothly varying SFHs.

Fig.~\ref{fig:mtime} and \ref{fig:mage} show that the success of the mass recovery does not depend on the time elapsed in the simulation or (connected to it) mass-weighted age of the stellar population. Hence, the stage of the merger and the resulting complication of the non-uniform geometry and dust attenuation do not play a role in deriving global properties of galaxies.

It is clearly shown in Fig.~\ref{fig:magediff} that mismatches between the true and inferred mass-weighted stellar age are the key determinant of  the accuracy of the mass recovery. For almost all the simulated galaxies for which the age is accurately recovered, so is the stellar mass. Moreover, for most of the models $\mbox{age}_{\rm SED}-\mbox{age}_{\rm simulations}$ is correlated with  $M_{\rm *\, SED}/M_{\rm *\, simulations}$.

Fig.~\ref{fig:mav} shows how the success of the stellar mass recovery depends on the optical dust attenuation. 
The $M_{\rm *\, SED}/M_{\rm *\, simulations}$ ratio recovered when using double SFHs does not depend on $A_V$.
This could  mean that either these models accurately account for the dust attenuation in the simulated {\smgs}, or that this parameter is not critical for recovering the stellar mass. Fig.~\ref{fig:mavdiff} shows that the latter possibility is correct --- even though most of the models cannot reproduce the $A_V$ accurately, they still result in the correct stellar masses, and there is no correlation between $A_{V{\rm, SED}}-A_{V{\rm, simulaitons}}$ and $M_{\rm *\, SED}/M_{\rm *\, simulations}$. This is because for the attenuation levels explored here, the effect of dust on the near-infrared emission, which critically influences the stellar mass determination, is minor. Only the C10~tau models exhibit an anti-correlation between $A_{V{\rm, SED}}-A_{V{\rm, simulaitons}}$ and $M_{\rm *\, SED}/M_{\rm *\, simulations}$, which is a consequence of the strong age-dust degeneracy (see Fig.~\ref{fig:avdiffagediff}) --- if too much dust is assumed, then the model adopts too low age resulting in the mass underestimation.

 In contrast, single-component SFHs underestimate the stellar masses of {\smgs} for high attenuations (Fig.~\ref{fig:mav}). This effect was found by \citet[][their Fig.~6, left]{lofaro13} for $z\sim1$ LIRGs and $z\sim2$ ULIRGs; they interpreted this result as an inability to recover a highly obscured stellar component with models that utilise only rest-frame optical and near-IR data. However, we do not see this effect for our double-burst SFH model, which does not utilise long-wavelength data either (C10~2c). 
Moreover, when we did not use long-wavelength data in {\sc Grasil} we obtained identical masses for our simulated {\smgs} (median $11.79$ instead of $11.73$).
Thus, we conclude that this issue is connected with the assumption of a too-simplistic SFH and therefore manifests itself only with single-burst models.

The analysis of the age and dust attenuation is summarized in Fig.~\ref{fig:avdiffagediff}. It shows the degeneracy in the SED fitting, as $\mbox{age}_{\rm SED}-\mbox{age}_{\rm simulations}$ correlates with $A_{V{\rm, SED}}-A_{V{\rm, simulaitons}}$.  It is instructive to point out that this correlation is not apparent for {\sc Magphys} (and very weakly for {\sc Grasil}), 
because 
$A_V$ is constrained by the energy-balance argument  in addition to the UV-optical colours of the attenuated stellar emission.  Fig.~\ref{fig:avdiffagediff} shows again that the age is more important in determining the stellar masses. The datapoints are colour-coded by the  $M_{\rm *\, SED}/M_{\rm *\, simulations}$ ratio, and when moving from  black/dark blue (mass underestimation), through light blue/green (correct mass) to orange/red (mass overestimation), the points clearly move from left (age underestimation) to right (age overestimation), but not at all from the bottom (dust underestimation) to the top (dust overestimation), which would be expected if accurate recovery of dust attenuation played an important role in the stellar mass determination.

We investigate the impact of  AGN emission on the stellar mass determination in Fig.~\ref{fig:agn}. Even though AGN can contribute as much as $\sim60$\% to the K-band luminosity, the ratio of the estimated and true stellar masses does not depend on the AGN fraction. We attribute this to the fact that the AGN emission becomes less significant at shorter wavelengths (away from the mid- and near-IR), and the SED codes correctly utilise the information at shorter wavelengths to recover the appropriate mass-to-light ratio.

The fact that despite the simplistic assumptions made in the SED modelling codes regarding the galaxy shape, dust distribution and attenuation curves, we were able to recover the correct stellar masses of the simulated {\smgs} gives confidence in SED modelling, even for galaxies as dusty as {\smgs}.
Moreover, the simulated {\smgs} analysed here reveal trends that are similar to those found for real {\smgs}: in \citet{michalowski12mass}, we demonstrated that the choice of the parametrization of the SFH may change the resulting stellar mass by a factor of $\sim2$--$3$. In particular, double SFHs result in correct stellar masses, whereas utilising a single-burst SFH leads to serious stellar mass underestimation.

The reason for the success of double SFHs is that they are a more general form that can accommodate both an older stellar population formed through quiescent star formation, which dominates the stellar mass formed in merging galaxies \citep[e.g.][]{hopkins10}, and a younger population formed in a recent starburst. The simulated {\smgs} studied in this work consist of both star-forming disc galaxies and merger-driven starbursts \citep{hayward11b, hayward12,hayward13}, and the double SFHs are able to recover the stellar masses accurately for both types. Consequently, for real  {\smgs}, the stellar masses inferred using double SFHs, which tend to be higher than those recovered using single SFHs, should be considered more accurate. 
As demonstrated previously \citep{michalowski12mass,koprowski14,koprowski15}, such higher stellar masses of {\smgs} imply that they are not significant outliers from the star-forming main sequence at $z\sim2$, 
and we show here that this is the case even if real {\smgs} are predominately major mergers, since the suite of simulated 
galaxies modelled here contains examples of both merging and isolated galaxies.

\section{Conclusions}
\label{sec:conclusion}

We have tested the reliability of stellar mass estimation using the SEDs of simulated {\smgs}. We found that the assumption of double SFHs leads to correct stellar masses on average,
whereas exponentially declining SFHs result in lower masses, but they are still consistent with the true values. Single-burst SFHs underestimate the stellar masses by $\sim0.2$ dex. We found that single-component SFHs underpredict stellar masses more significantly when the dust attenuation is greater, but this is not the case for double SFHs.
We also identified discrepancies between the true and inferred stellar ages (rather than the dust attenuation) as the key determinant of  the success of the mass recovery.
The scatter of $\sim0.3$ dex around the correct value is an intrinsic uncertainty in determining stellar masses from broad-band photometry through SED modelling; it originates from simplifying assumptions regarding the galaxy SFH, geometry, and attenuation.  Finally, we did not find evidence for a significant impact of AGN emission on the derived stellar masses, so notably, the reported differences are not due to whether or not the AGN emission has been subtracted before fitting for stellar masses. 

Thus, we conclude that studies based on the higher masses inferred from fitting the SEDs of real {\smgs} with 
double SFHs are most likely to be correct, implying (as shown, for example, by \citealt{michalowski12mass}) that  
{\smgs} lie on the high-mass end of the main sequence of star-forming galaxies. This conclusion 
appears robust to assumptions of whether or not {\smgs} are driven by major mergers, since the suite of simulated 
galaxies modelled here contains examples of both merging and isolated galaxies.


\begin{acknowledgements}

We thank Joanna Baradziej, Desika Narayana, Dan Smith, and our anonymous referee for useful suggestions.
MJM acknowledges the support of the Science and Technology Facilities Council, SUPA Postdoctoral and Early Career Researcher Exchange Program,
and the hospitality at the Harvard-Smithsonian Center for Astrophysics. 
CCH is grateful to the Klaus Tschira Foundation for financial support and acknowledges the hospitality of the Aspen Center for Physics, which is supported by the National Science Foundation Grant No. PHY-1066293.
JSD acknowledges the  support of the Royal Society via a Wolfson Research Merit award, the support of the European Research Council  through an Advanced Grant and the contribution of the EC FP7 SPACE project ASTRODEEP (Ref.No: 312725). 
VAB, MC and FC acknowledge the support of the Science and Technology Facilities Council.
This research has made use of  
the NASA's Astrophysics Data System Bibliographic Services.

\end{acknowledgements}



\end{document}